\documentclass[a4paper]{article}
\usepackage{import}
\usepackage{pgfplots}
\usepackage{multicol}
\usepackage{multirow}
\usepackage{xcolor}
\usepackage{makecell}
\usepackage{INTERSPEECH2019}

\title{MIRNet: Learning multiple identities representations in overlapped speech}
\name{Hyewon Han, Soo-Whan Chung and Hong-Goo Kang}

\address{
  Department of Electrical \& Electronic Engineering, Yonsei University, Seoul, South Korea}
\email{[hwhan, jsh6293]@dsp.yonsei.ac.kr, hgkang@yonsei.ac.kr}
\begin{document}

\maketitle

\begin{abstract}
Many approaches can derive information about a single speaker's identity from the speech by learning to recognize consistent characteristics of acoustic parameters.
However, it is challenging to determine identity information when there are multiple concurrent speakers in a given signal.
In this paper, we propose a novel deep speaker representation strategy that can reliably extract multiple speaker identities from an overlapped speech.
We design a network that can extract a high-level embedding that contains information about each speaker's identity from a given mixture.
Unlike conventional approaches that need reference acoustic features for training, our proposed algorithm only requires the speaker identity labels of the overlapped speech segments.
We demonstrate the effectiveness and usefulness of our algorithm in a speaker verification task and a speech separation system conditioned on the target speaker embeddings obtained through the proposed method.

\end{abstract}

\section{Introduction}
\label{sec:intro}

Speech is one of the most widely used media for recognizing peoples' identities due to its distinctive characteristics.
Typically, an individual's vocal identity is represented by acoustic characteristics such as pitch, formant, and speaking styles that are consistently observed in speech signals~\cite{atal1972automatic,wolf1972efficient,liu1990study}.
In early studies, speaker representations were obtained using statistical models such as Gaussian mixture models~\cite{reynolds1995GMM} with Joint Factor Analysis or Support Vector Machines~\cite{wan2000SVM,campbell2006support, dehak2010ivector}.
By removing rapidly varying acoustic features caused by linguistic changes, these methods obtained normalized speaker-discriminative features with clustering or classification methods~\cite{soong1987VQ}.

Thanks to advances in deep learning and the availability of large-scale datasets~\cite{mclaren2016speakers,chung2018voxceleb2}, deep neural network-based speaker modeling strategies have recently shown great success in speaker recognition.
For example, in~\cite{variani2014dvector,snyder2018xvector}, neural network models were trained to perform classification given the reference speaker labels, from which representations for speaker information were obtained from the output of the last hidden layer. These methods showed better performance than previous statistical approaches such as i-vectors~\cite{dehak2010ivector}.

Speaker identity information obtained from such methods can be applied to a variety of speech interface related applications.
For example, the accuracy of automatic speech recognition (ASR) systems can be improved by using speaker identity information to reduce the bias caused by speaker-dependent characteristics~\cite{tan2016speaker,pironkov2016speaker}.
The performance of tasks such as speech enhancement and separation can also be improved when this information is available, which can be used to specify or represent the desired voice components in input signals distorted by noise, reverberation or interfering speech~\cite{delcroix2018speakerbeam,Wang2019VoiceFilter,xu2019time}.

However, the aforementioned approaches are often inconvenient for real applications because target speakers must first enroll their information before post-tasks can be performed.
Therefore, they cannot be used with any unenrolled speakers. 
It is possible to directly extract speaker representations from a mixed input signal, but these representations are prone to error; thus, the overall performance of the post-task can degrade as a result of incorrect guidance on speaker identity.

In this paper, we propose a novel deep learning-based speaker representation method that extracts multiple identities in the case of simultaneous speech. Our proposed method includes two contributions: a speaker model and its training strategy.
First, we construct a speaker model with three modules: {\em speech analysis}, {\em spectral attention}, and {\em speaker embedding}. The speech analysis module transforms the input spectrum into a latent domain and the spectral attention module estimates each speaker's spectral information in the latent space using a temporal attention layer, after which the speaker embedding module finds a representation for each speaker's identity. Using this structure, the network is able to directly estimate speaker identities from mixed speech.
Second, we propose an effective training strategy for this network that only uses speaker identity information. 
It utilizes a decision method for speaker permutations to alleviate the ambiguity of speaker assignment.
Our approach is novel in that it models speaker identities from the mixed speech with a single network without requiring any clean spectral information. Using this method, we can acquire each speaker's identity information even in cases in which clean utterances are not enrolled. 

To verify the effectiveness of our proposed approach, we perform several experiments that utilize the extracted identity information.
We first measure speaker verification performance in an overlapped speech scenario.
By measuring the similarity between obtained speaker identities from pairs of mixed speech, we show that our network effectively models speaker identity information. 
We also show that our extracted speaker representations from mixed speech contain distinctive speaker information by applying them to a speaker-conditioned speech separation task.

The rest of the paper is organized as follows. Section~\ref{sec:related} describes prior research that is related to our work. The proposed network structure and training strategy for learning speaker identity are shown in Section~\ref{sec:proposed}. Experimental results and applications for our proposed method are shown in Section~\ref{sec:experiment}, with final conclusions in Section~\ref{sec:conclusion}.

\begin{figure*}[t]
    \centering
    \begin{minipage}[b]{0.510898\linewidth}
        \includegraphics[width=\linewidth]{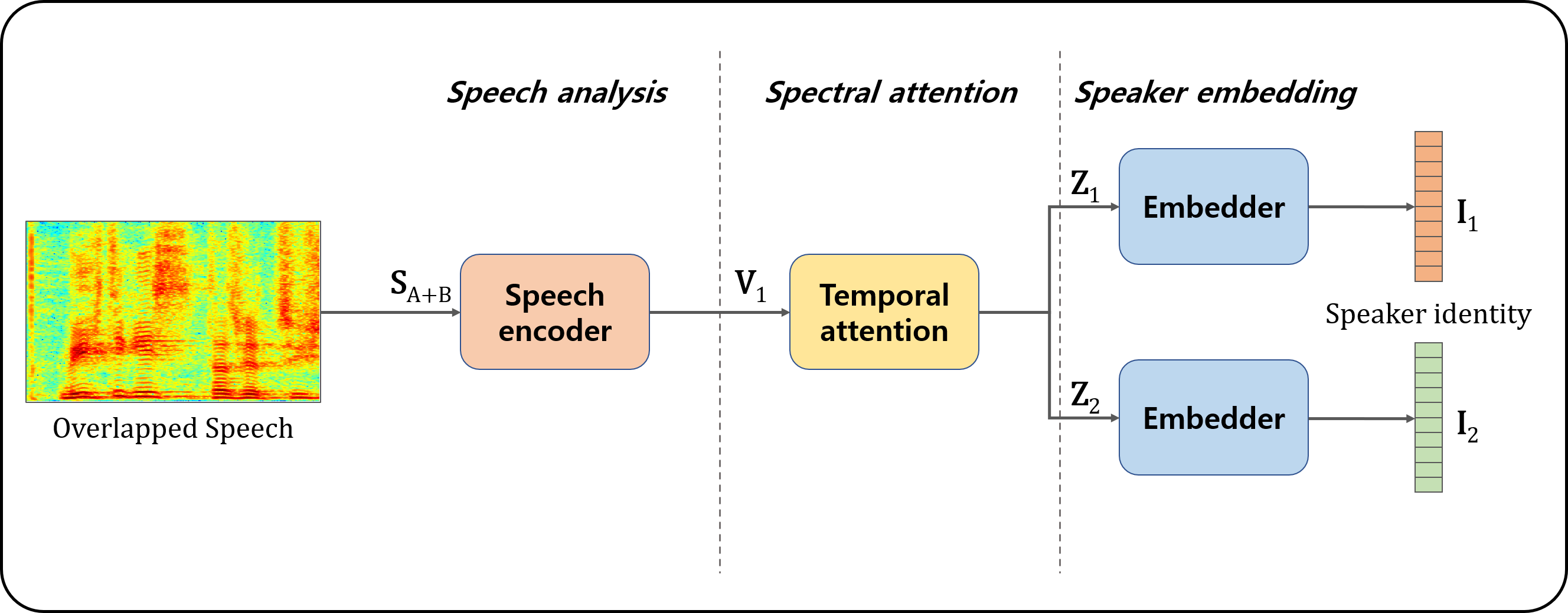}
        \centerline{(a) Overall structure of MIRNet}
    \end{minipage}
    \hfill
    \begin{minipage}[b]{0.459102\linewidth}
        \includegraphics[width=\linewidth]{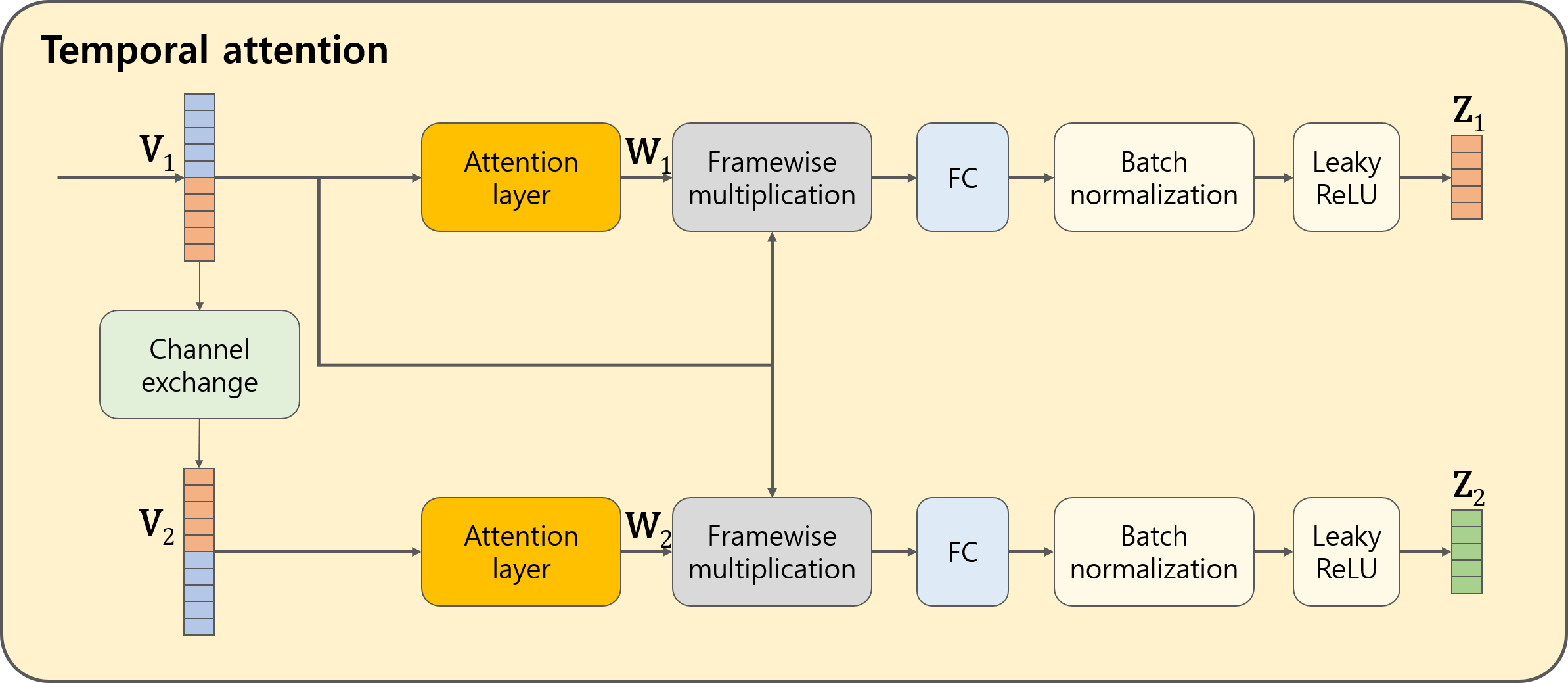}
        \centerline{(b) Temporal attention network}
    \end{minipage}
    \caption{Demonstration of the proposed speaker representation method.}
    \vspace{-10pt}
    \label{fig:blockdiagram}
\end{figure*}

\section{Related Work}
\label{sec:related}
\subsection{Speaker identity representation}
Speaker identity representation is the task of deriving representative embeddings for this information by mapping acoustic features into a latent space.
Various deep learning-based approaches have been proposed to improve the performance of speaker representations.
D-vectors~\cite{variani2014dvector} are a high level speaker representation embedding extracted from the last hidden layer of a deep neural network architecture that is designed to perform a speaker identification task.
To capture the sequential information of speech signals, x-vectors \cite{snyder2018xvector} exploit a time-delay neural network~(TDNN) architecture that consists of frame-wise outputs and statistical pooling at the sentence level. 
Also, various methods are proposed to model effective speaker representation by modifying the aggregation method~\cite{tang2019deep,MSA}.

Various training criteria have been investigated to further improve the uniformity of output speaker representations.
The most well-known loss function is cross-entropy loss, which facilitates the categorization of speaker identity. 
Recently, metric-based learning criteria such as triplet loss~\cite{zhang2017end,li2017deep} and prototypical loss~\cite{wang2019centroid,kye2020meta} have been used to enlarge the similarity between same speaker pairs but minimize the similarity between different speaker pairs.
In this work, we apply cross-entropy loss for a learning network to categorize the speaker identities.

\subsection{Permutation invariant training}
Speech separation is the task of estimating individual signals for each speaker from a mixed speech signal. 
One of the most difficult issues in this work is how to avoid the speaker permutation problem, {\em i.e.} how to correctly assign the speaker ID of the separated signal in each processing frame.
To alleviate this problem, the permutation invariant training~(PIT) criterion was proposed~\cite{Yu2017PIT,Kolbaek2017uPIT}, which simply considers all of the loss terms by calculating every possible permutation of candidate pairs. From those candidates, the network determines optimal speaker pairs that minimize the error. Inspired by this approach, our training criterion also assigns a loss function for all possible pairs and finds the optimal assignment. 
However, we compute the PIT loss on estimated speaker identity information such as embeddings or the distribution of speaker classes, not the separated spectral bins as the conventional method did.

\section{Multiple identities representation network (MIRNet)}
\label{sec:proposed}

Although speaker information is mingled in an overlapped speech signal, it is possible to extract a target speaker's identity if an appropriate target-related condition vector is given~\cite{shi2020demix}.
Conventionally, a reference speech signal is used as a condition vector, which necessitates a cumbersome pre-enrollment process.
In this work, we separate speaker identity information in an overlapped signal without providing any reference signal.
Figure~\ref{fig:blockdiagram} illustrates a block diagram of the proposed multiple identities representation network (MIRNet).
Although the MIRNet can be generalized to an arbitrary number of speaker inputs, we fix the number of speakers to two for simplicity in this paper.
The proposed network consists of three stages: speech analysis, spectral attention and speaker embedding.

\subsection{Speech analysis}
In the speech analysis stage, an input speech signal is transformed into a latent domain representation, $\mathbf{V}_1$.
\vspace{-2pt}
\begin{equation}
    \mathbf{V}_1=E_{S}(S_{A+B}), 
\end{equation}
where $E_{S}$ is a speech encoder network and $\mathbf{V}_1$ is an embedding with $2D$-channels.
The input to the encoder, $S_{A+B}$, denotes a linear magnitude spectrum on a logarithm scale.
We do not use the mel-spectrum that is popularly used in speaker embedding tasks because of its over-smoothed spectral characteristics, which makes it difficult to distinguish one speaker from another. 
We construct the architecture of the speech encoder based on 1-D convolution layers. The detailed parameter settings of the speech encoder are described in Table~\ref{tab:network}.
The speech encoder emits spectral embeddings with $2D$ channels, where we assume that each $D$ channel contains the information for one speaker.

\subsection{Spectral attention}
In the spectral attention stage, we extract two different sets of embeddings from the spectral embedding output of the speech encoder using an attention mechanism.
The rationale for the use of an attention mechanism is as follows. 
Some frames that are fully-overlapped with two speakers do not represent speakers' identities well, but other frames clearly represent each speaker's identity when they are spoken by only a single speaker. In addition, some frames contain only silence.
We compute frame-wise attention weights from the spectral embeddings to estimate the amount of importance that each frame implies.

Figure~\ref{fig:blockdiagram} illustrates a detailed block diagram of the self-attention mechanism~\cite{lin2017selfatt}.
We first obtain two embedding vectors, one from the spectral embedding $\mathbf{V}_1$ and the other from $\mathbf{V}_2$, re-structured by a channel exchange.
Since the attention layer shares parameters for attention vectors, half of the channels of $\mathbf{V}_1$ are flipped onto the channel axis to create a different input embedding $\mathbf{V}_2$, which is used to extract another attention vector $W_2$.
\vspace{-2pt}
\begin{equation}
\mathbf{V}_2=Concat(\mathbf{V}_{1, D+1:2D}, \mathbf{V}_{1, 1:D}).
\end{equation}
The attention vectors are related to the framewise power of each speaker, which refers to the speaker information in each frame.
\begin{figure}[t]
  \centering
  \begin{minipage}[b]{\columnwidth}
    \includegraphics[width=\linewidth]{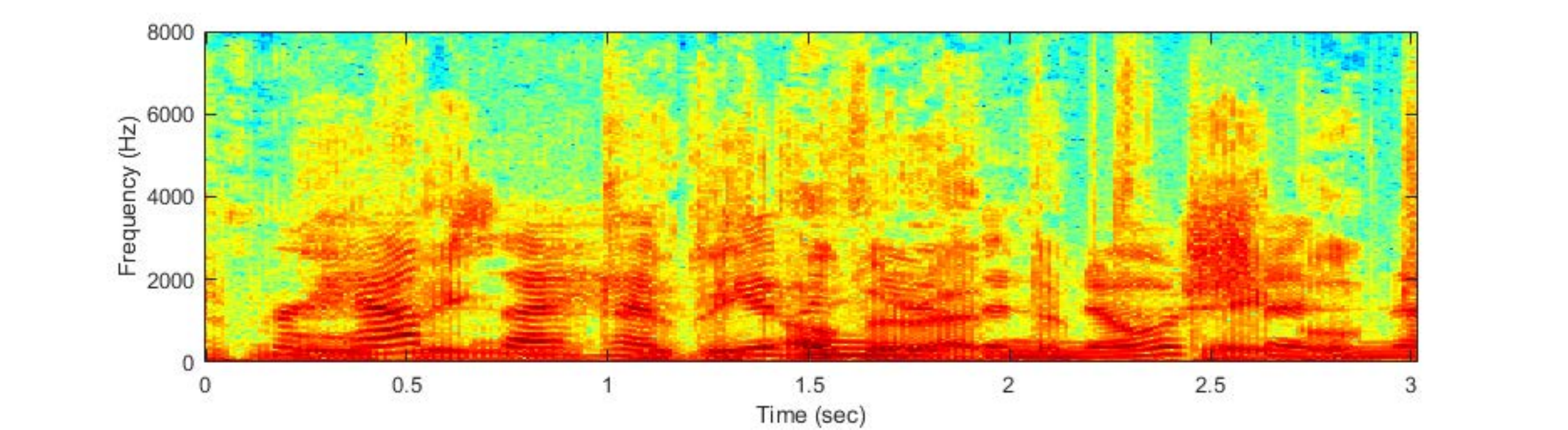}
  \end{minipage}
  \begin{minipage}[b]{\columnwidth}
    \includegraphics[width=\linewidth]{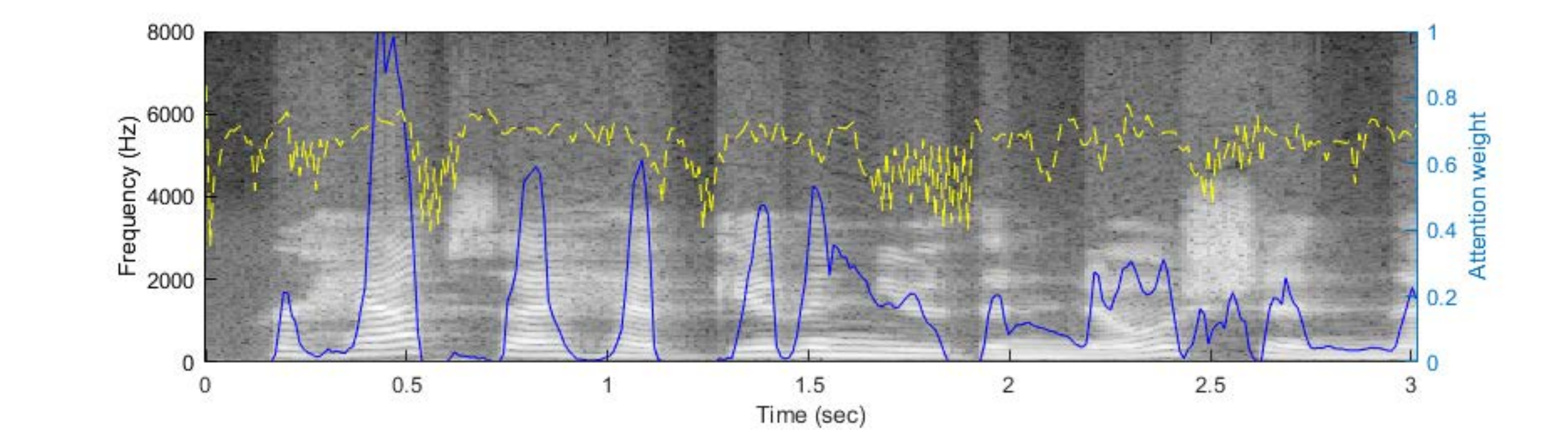}
  \end{minipage}
  \begin{minipage}[b]{\columnwidth}
    \includegraphics[width=\linewidth]{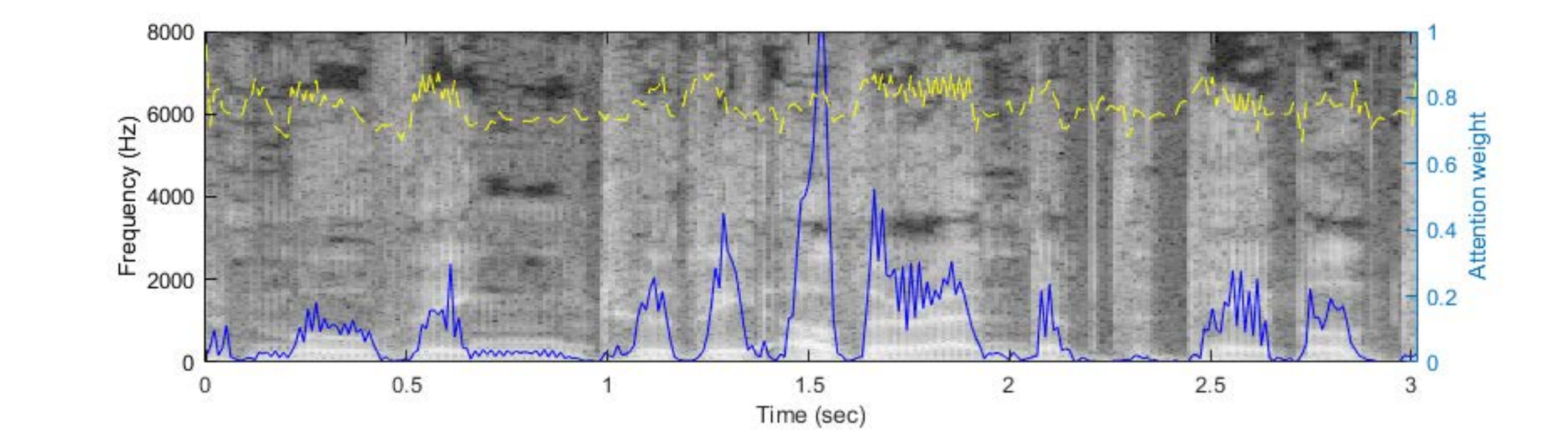}
  \end{minipage}
  \caption{Visualization of analysis on temporal attention for each speaker in a mixture. {\bf Top}: Spectrogram of input speech signal $S_{A+B}$. {\bf Mid}: Spectrogram of speech $S_A$. {\bf Bottom}: Spectrogram of speech $S_B$. Yellow line shows attention weight values, and the blue line is the power contour of speech signal.}
  \vspace{-10pt}
  \label{fig:attention}
\end{figure}
They are each multiplied with the spectral embedding $\mathbf{V}_1$ and used as inputs to the following fully-connected layer with a non-linear activation function. 
From the spectral attentive stage, the network outputs $\mathbf{Z}_1$ and $\mathbf{Z}_2$ where they contain discriminative information of each speaker.

Figure~\ref{fig:attention} illustrates an example log magnitude spectrogram of mixed speech from two speakers and those for each speaker individually, as well as power contours and attention weights estimated by the self-attention mechanism.

\subsection{Speaker embedding}
The speaker embedding network extracts speaker vectors from the outputs of the spectral attention stage.
The network architecture is similar to the one used for speaker verification, where we use ResNet-18~\cite{szegedy2017inception} as a backbone network while changing the pooling strategy to be temporal average pooling~(TAP).

\subsection{Training strategy}
For training, we randomly mix signals from two different speakers.
The networks are jointly trained using cross-entropy loss with a classifier for the speaker embeddings $I_1$ and $I_2$.
The permutation invariant training method is used to solve the permutation problem frequently occurring in speech separation tasks.
The entire training criterion is as follows:
\vspace{-2pt}
\begin{equation}
\mathcal{L}=min~(\mathcal{L}_1, \mathcal{L}_2),
\end{equation}
\begin{equation}
\mathcal{L}_1=\mathcal{L}_{CE} (y_1, y_{A}) + \mathcal{L}_{CE} (y_2, y_{B}),
\end{equation}
\begin{equation}
\mathcal{L}_2=\mathcal{L}_{CE} (y_2, y_{A}) + \mathcal{L}_{CE} (y_1, y_{B}),
\end{equation}
where $\mathcal{L}_{CE}$ is cross-entropy loss.

Suppose that we know the labels of speaker $y_A, y_B$ when we generate the mixture $S_{A+B}$, and the classifier estimates speaker identities ($y_1$, $y_2$).
Then, the losses between the labels $y_A, y_B$ and $y_1, y_2$ are computed for every pair to find the optimal speaker pairs.

\section{Experiments}
\label{sec:experiment}
In this section, we describe two experiments on a speaker verification task and a speaker-conditioned speech separation task to prove the effectiveness of our method.

\subsection{Speaker verification}
\label{subsec:speaker_verification}
\subsubsection{Dataset and neural network settings}
The speaker representation network was trained with the LibriSpeech corpus~\cite{panayotov2015librispeech}, which contains 2,238 speakers.
We randomly selected two speech samples from all the different pairs of speakers to make mixed input signals in every epoch.
In the training stage, 3 second segments chosen at random offsets were used as inputs to the network. 
In each epoch, the model was trained using 93 hours of mixed signals for training and 15 hours of mixed signals for validation.
The input log-scaled spectrum was calculated every 10 ms with an analysis frame length of 32 ms. The FFT size was set to 512; thus, the input dimension was 257.
To measure the speaker embedding performance of the network in speaker verification, 200 evaluation pairs were generated for each acceptance and rejection scenario using the \(\textit{dev-clean}\), \(\textit{dev-others}\), \(\textit{test-clean}\) and \(\textit{test-others}\) subsets which contain a total of 146 speakers. 

\begin{table}[t]
\centering
\caption{Details of model parameter settings for speech encoder and attention layer}
\vspace{-10pt}
\label{tab:network}
\begin{center}
\begin{tabular}{c|c|c|c}
\toprule
	\multicolumn{4}{c}{\textbf{Speech encoder}} \\ \midrule
	\bf Layer &	\bf Non-linearity & \bf Channels & \bf Kernel\\ \midrule \midrule
    conv1 & \multirow{6}{*}{\makecell{Leaky\\ReLU\\($\alpha=0.2$)}} & {512} & {5}\\ 
    conv2 &                                         & {512} & {3}               \\ 
    conv3 &                                         & {512} & {3}               \\ 
    conv4 &                                         & {512} & {1}               \\ 
    conv5 &                                         & {1,500} & {1}             \\ 
    conv6 &                                         & {514} & {1}\vspace{3pt}   \\ \toprule
    \multicolumn{4}{c}{\textbf{Attention layer}} \\ \midrule
    \bf Layer &	\bf Non-linearity   & \multicolumn{2}{c}{\bf Channels}   \\ \midrule\midrule
    fc1 & Tanh                      & \multicolumn{2}{c}{64}            \\ \midrule
    fc2 & Sigmoid                   & \multicolumn{2}{c}{1}             \\
\bottomrule
\end{tabular}
\end{center}
\vspace{-10pt}
\end{table}

The detailed network architecture is summarized in Table~\ref{tab:network}.
We use a 257-dimensional log-spectrum as the input, and the output channel dimension of the last layer in the speech encoder is 514, two times that of the input spectrum dimension. For the attention layer, the sigmoid activation function is used to limit the range of the attention weights from 0 to 1. The obtained attention vector is mapped into a 257-dimensional vector using a fully-connected layer. 

\subsubsection{Evaluation protocol}
To prove whether the proposed model represents discriminative characteristics of speaker identities well, we measure the equal error rate~(EER) for the speaker verification task using the extracted embeddings from the input mixed speech.
While conventional speaker verification methods use positive and negative embedding pairs for evaluation, our method needs a new protocol setup since the output embeddings are permuted and not assigned to speaker labels.

Our evaluation method considers two scenarios to measure EER: acceptance and rejection scenarios.
We prepare three mixtures $S_{A+B}$, $S_{A'+C}$ and $S_{C'+D}$, and each speech segment provides speaker identities ($I_{A}, I_{B}$), ($I_{A'}, I_{C}$) and ($I_{C'}, I_{D}$), respectively.
While ($I_{A}, I_{B}$) is set as the anchor embedding pair, we use the positive and negative samples from ($I_{A'}, I_{C}$) and ($I_{C'}, I_{D}$).
To compute EER, $d_p$ and $d_n$ are the distances for the acceptance and the rejection, with equations as shown below:
\begin{equation}
    d_p=\min_{i,j}(d(I_{i|{A+B}}, I_{j|{{A'}+C}}))
    \label{eq:pos}
\end{equation}
\begin{equation}
    d_n=\min_{i,j}(d(I_{i|{A+B}}, I_{j|{{C'}+D}}))
    \label{eq:neg}
\end{equation}
where $i,j\in\{1,2\}$.
The distances are computed from the closest identity pairs between the anchor and positive and between the anchor and negative samples using Euclidean distance, denoted as $d(\cdot)$. $I_{i|{A+B}}$ denotes the identity from the $i$-th output from mixture $S_{A+B}$.

\subsubsection{Experimental results}
We obtained an EER of 5.00$\%$ and 6.00$\%$ with the aforementioned evaluation method for the seen and unseen speaker cases, respectively. From this result, we can see that our trained network successfully represents speaker identities from a mixed input. This result also demonstrates the efficiency of the decision method for avoiding the permutation problem.

\begin{figure}[t]
  \centering
  \begin{minipage}[b]{0.45\columnwidth}
    \includegraphics[width=\linewidth]{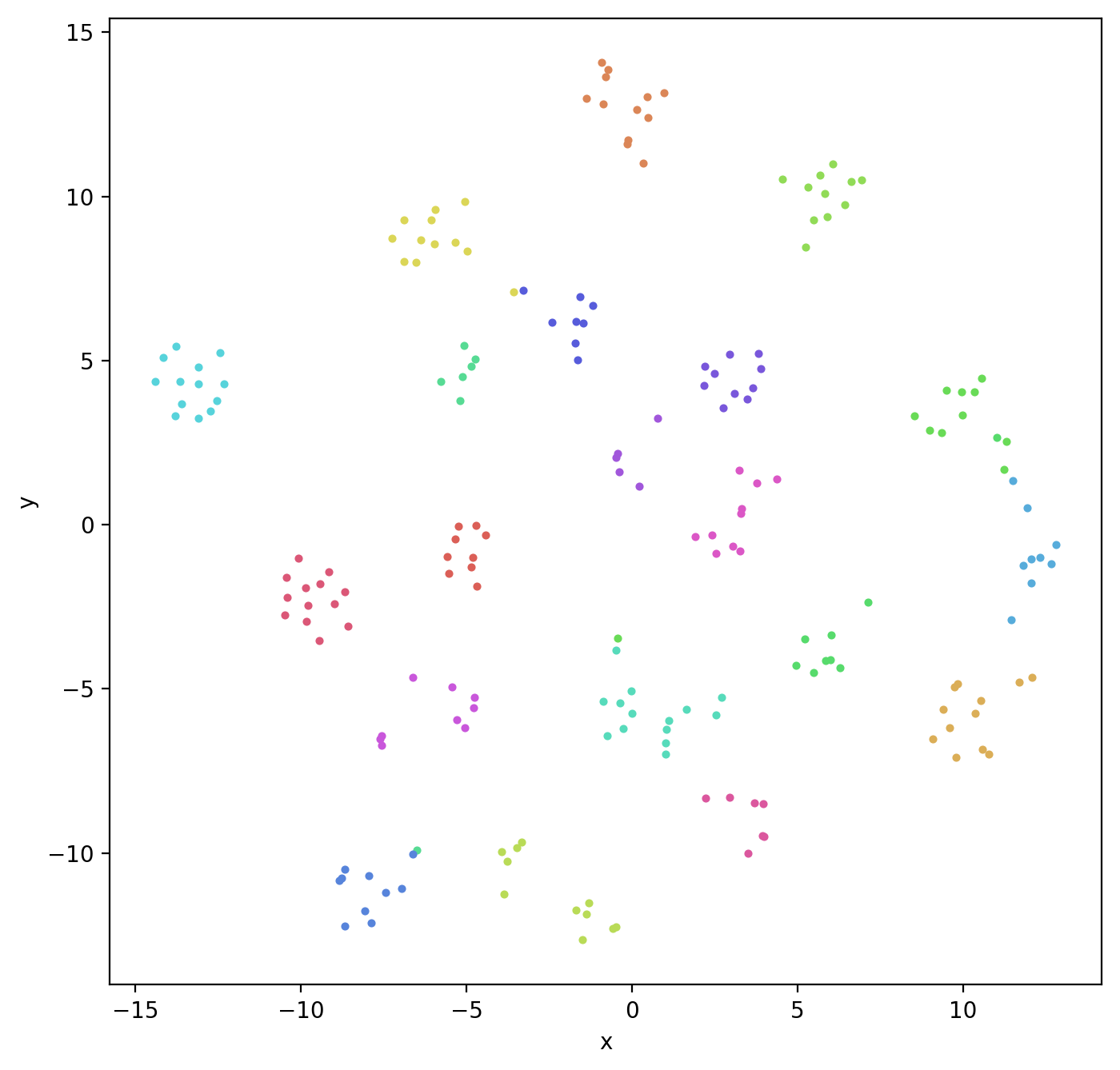}
    \centerline{(a) Speaker embedding}\medskip
  \end{minipage}
  \begin{minipage}[b]{0.45\columnwidth}
    \includegraphics[width=\linewidth]{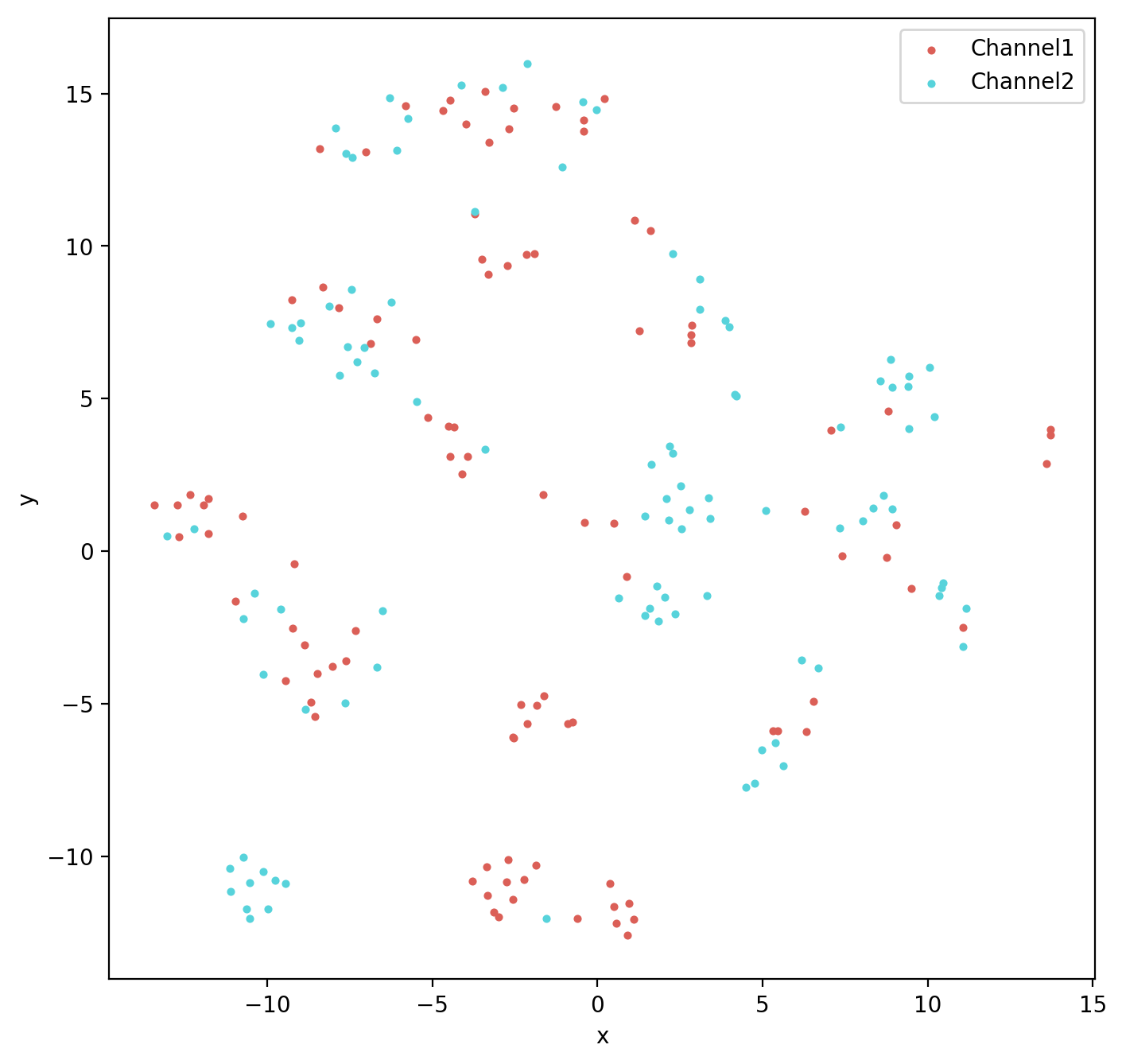}
    \centerline{(b) Embedding of each channel}\medskip
  \end{minipage}

  \caption{t-SNE visualization of extracted speaker embedding from 20 random speakers. (a) represents speaker embeddings labeled with our permutation decision method. (b) represents the speaker embeddings from the each channel.}
\label{fig:tSNE}
\vspace{-5pt}
\end{figure}

Figure~\ref{fig:tSNE} depicts a t-SNE visualization of speaker embeddings extracted by the proposed network architecture.
The speech samples are mixed with combinations generated from 20 different speakers, and speaker assignment is decided with our proposed permutation decision method.
We can see that embeddings from the same speaker are clustered closely together.
We also checked the variances of the channels containing speaker identity information to confirm that they effectively model the information without biased channel values.
Therefore, we also visualized embeddings extracted from each channel

\subsection{Speaker-conditioned speech separation}
\subsubsection{Task description}
Speaker-conditioned speech separation is the task of verifying whether speaker information embeddings are informative enough for speaker separation in overlapped speech signals.
In~\cite{Wang2019VoiceFilter}, VoiceFilter was proposed to improve speech separation performance using speaker identities extracted from pre-enrolled reference speech.
It dramatically improved the separation performance, but it was not applicable to speech signals spoken by speakers that had never been seen before.

Here, we extract multiple identities from overlapped speech first and separate the target speakers' voices from the signal.
We use VoiceFilter as our baseline for comparison, which utilizes an independent speaker model. We also compare with a separation system without speaker conditioning using uPIT~\cite{Kolbaek2017uPIT}.

\subsubsection{Implementation details}
\noindent\textbf{Dataset.}
The training dataset for our speaker model is identical to the one described in Section~\ref{subsec:speaker_verification}.
The speaker model for VoiceFilter is pre-trained using the VoxCeleb2 dataset~\cite{chung2018voxceleb2}.
We use the WSJ0-2mix dataset~\cite{Hershey2016DeepClustering} to train and evaluate speech separation models for the baseline and our method.
Speech signals in WSJ0-2mix are sliced every 10ms with 32ms frame length, from which they are transformed into the time-frequency domain using the Fourier transform.

\noindent\textbf{Model setup.}
We use a model pre-trained on VoxCeleb2 whose performance is 2.23\% EER~\cite{chung2020in} as the baseline speaker model, which consists of 34 convolution layers with Inception.
Model parameters for the speech separation model are the same as those used in~\cite{Wang2019VoiceFilter}. For the model without speaker conditioning, we use a bidirectional LSTM (BLSTM) structure identical to VoiceFilter while doubling the dimension of the fully-connected layer and output for separate speech.

\noindent\textbf{Evaluation metric.}
For objective evaluation, we measure signal-to-distortion ratio improvement~(SDRi), which is typically used to represent speech separation performance.
We also calculate the perceptual evaluation of speech quality improvement~(PESQi), which quantifies the perceptual scoring for separated speech signals.

\subsubsection{Separation results}
Table~\ref{tab:Voicefilter} shows results comparing our method with the baseline.
Our separation method results in slightly lower scores than the baseline model.
Nevertheless, it is noteworthy that our method is able to achieve reliable performance on this task even with unseen speakers.
It should be noted that the baseline uses speaker identities extracted from enrolled reference speech; this not only gives it a significant advantage in the speech separation task, but also means that it cannot be used at all without pre-enrolling speakers.
In addition, since MIRNet was trained using a much smaller dataset compared to the baseline model, there should still be margins for it to improve.


\begin{table}[t]
\centering
\caption{Performance on the speech separation task}
\vspace{-10pt}
\label{tab:Voicefilter}
\begin{center}
\begin{tabular}{c|c|c}
\toprule
\bf Speaker encoder	& \bf SDRi~(dB) & \bf PESQi \\  
\midrule\midrule
Chung et al.~\cite{chung2020in} & 6.147 & 0.480 \\ \midrule
\bf Proposed & 6.097 & 0.376 \\ \midrule
uPIT & 5.897 & 0.367 \\ \bottomrule
\end{tabular}
\end{center}
\end{table}
\vspace{-2pt}

\section{Conclusion}
\label{sec:conclusion}
In this work, we proposed a novel method to estimate latent embeddings from overlapped speech that reliably represent speaker identity information. 
The proposed network consists of speech analysis, spectral attention, and speaker embedding stages which extract information on multiple speaker identities.
To make the network learn this information while finding an optimal assignment, we proposed a speaker identity decision procedure based on permutation invariant training.
Experimental results showed that the proposed network can derive individual speakers' identity information from the mixtures without using acoustic information extracted from reference speech signals.
In addition, the resulting embeddings showed reliable performance on a speaker-conditioned speech separation task, where it has an advantage in that it can be applied even to speakers for which clean reference speech is unavailable.

\vspace{3pt}
\noindent\textbf{Acknowledgements.} 
This research was sponsored by Naver Corporation.

\clearpage

\bibliographystyle{IEEEtran}

\bibliography{temp/0_main}

\end{document}